\documentclass[conference,final,a4paper]{IEEEtran}

\usepackage{amssymb}
\usepackage{graphicx}

\usepackage{url}
\usepackage[utf8]{inputenc}
\usepackage{gensymb}
\usepackage{listings}
\usepackage{color}
\usepackage[ruled,linesnumbered]{algorithm2e}
\usepackage{cite}
\usepackage{verbatim}
\usepackage{comment}
\usepackage{amsmath,amssymb,amsfonts}
\usepackage{algorithmic}
\usepackage{textcomp}
\usepackage{tabularx}
\usepackage{xcolor}
\usepackage{multirow}
\usepackage[export]{adjustbox}
\usepackage[acronym]{glossaries}
\usepackage[justification=centering]{caption}
\def\BibTeX{{\rm B\kern-.05em{\sc i\kern-.025em b}\kern-.08em
    T\kern-.1667em\lower.7ex\hbox{E}\kern-.125emX}}

\newacronym{ders}{DERs}{distributed energy resources}
\newacronym{ml}{ML}{machine learning}
\newacronym{dr}{DR}{demand response}
\newacronym{iot}{IoT}{Internet of Things}
\newacronym{hil}{HIL}{hardware-in-the-loop}
\newacronym{ghg}{GHG}{greenhouse gas}
\newacronym{evs}{EVs}{electric vehicles}
\newacronym{admm}{ADMM}{Alternating Direction Method of Multipliers}
\newacronym{dsm}{DSM}{demand side management}
\newacronym{pv}{PV}{photovoltaics}
\newacronym{w}{W}{watt}
\newacronym{kw}{kW}{kilowatt}
\newacronym{cfl}{CFL}{compact fluorescent light}
\newacronym{lolp}{LOLP}{loss of load probability}
\newacronym{hvac}{HVAC}{heating, ventilation, and air conditioning}
\newacronym{ev}{EV}{electric vehicle}
\newacronym{om}{O\&M}{operations and maintenance}
\newacronym{idf}{IDF}{input data file}
\newacronym{doe}{DOE}{U.S. Department of Energy}
\newacronym{sce}{SCE}{Southern California Edison}
\newacronym{tou}{TOU}{time-of-use}

\makeatletter  
\def\ps@IEEEtitlepagestyle{%
  \def\@oddfoot{\mycopyrightnotice}%
  \def\@evenfoot{}%
}
\def\mycopyrightnotice{%
  {\footnotesize 978-1-6654-4421-7/21/\$31.00 \textcopyright2023 IEEE\hfill}
  \gdef\mycopyrightnotice{}
}

\makeatother  
\begin{document}


 \title{\Large Strategizing EV Charging and Renewable Integration in Texas
\vspace{-2ex}}

\author{
\IEEEauthorblockN{Mohammad Mohammadi}
\IEEEauthorblockA{Grid Fruit, LLC \\
Austin, TX, USA \\
mmohamm2@alumni.cmu.edu\vspace{-3ex}}

\and

\IEEEauthorblockN{Jesse Thornburg}
\IEEEauthorblockA{Carnegie Mellon University Africa \\
Grid Fruit, LLC \\
jesse@gridfruit.com\vspace{-3ex}}

}
\maketitle

\begin{abstract}

Exploring the convergence of electric vehicles (EVs), renewable energy, and smart grid technologies in the context of Texas, this study addresses challenges hindering the widespread adoption of EVs. Acknowledging their environmental benefits, the research focuses on grid stability concerns, uncoordinated charging patterns, and the complicated relationship between EVs and renewable energy sources. Dynamic time warping (DTW) clustering and k-means clustering methodologies categorize days based on total load and net load, offering nuanced insights into daily electricity consumption and renewable energy generation patterns. By establishing optimal charging and vehicle-to-grid (V2G) windows tailored to specific load characteristics, the study provides a sophisticated methodology for strategic decision-making in energy consumption and renewable integration. The findings contribute to the ongoing discourse on achieving a sustainable and resilient energy future through the seamless integration of EVs into smart grids.

\end{abstract}
\vspace{0.05in}
\begin{IEEEkeywords}
Electric vehicles, Smart Grid
\end{IEEEkeywords}

\section{Introduction}

Grid-connected transportation systems have long been considered a promising solution for curbing greenhouse gas and pollutant emissions. Among these solutions, Battery Electric Vehicles (BEVs) emerge as particularly compelling. However, despite their environmental advantages, formidable technological barriers, such as low battery capacity, high costs, and the absence of a widespread charging infrastructure, continue to impede the widespread adoption of plug-in vehicles.\cite{donateo2015evaluation}.
The positive impact of electric vehicles on urban pollution is evident as emissions shift from urban centers to fossil fuel chimneys typically located outside cities. Additionally, power plant emissions are more effectively regulated compared to vehicle tailpipes, further contributing to the appeal of electric vehicles \cite{sioshansi2010cost, moura2023maximizing}.

The surge in Electric Vehicles (EVs) is not only fueled by environmental consciousness but is also buoyed by government incentives, exemplified by measures like the Inflation Reduction Act in the US. A global emphasis on sustainability and the reduction of greenhouse gas emissions further propels the rise of EVs.\cite{Mohammadi2023, moura2021multi}
Recently, many countries have implemented policies, primarily in the form of tax incentives, aimed at bolstering the number of electric vehicles in their fleets. These policies seek to diminish pollutant emissions from traffic and enhance air quality, especially in urban areas where the population is significantly exposed to air pollutants, and traffic remains a major emission source.\cite{ferrero2016impact}

The repercussions of traffic emissions extend beyond immediate air quality issues, encompassing indirect human health effects resulting from rising atmospheric greenhouse gases and climate-related changes. Anticipated impacts include heightened public health risks due to climate change, extreme weather events, wildfires, and fluctuations in both indoor and ambient air pollutant levels.

While increasing the market share of EVs is a pivotal step toward achieving environmental goals, the actual benefits hinge on numerous factors, such as the energy pathway, energy generation profile, types of air pollutants and greenhouse gases, and the specific type of EV. Notably, EVs do not universally possess the potential to minimize all particulates and greenhouse gas emissions.\cite{requia2018clean}
Studies, such as those conducted in the U.S. state of Texas by Nichols et al. (2015), highlight the nuanced impact of EVs. While they can reduce GHG emissions, NOx, and PM10, they may generate significantly higher emissions of SO2 compared to Internal Combustion Engine (ICE) vehicles \cite{nichols2015air}.

Existing researches predominantly focuses on emissions reduction through the introduction of different fuels \cite{zhang2014can} or the implementation of new urban traffic policies \cite{cheng2015urban}. Enrico's 2015 study, for instance, assumes that electric vehicles replace all vehicles, estimating a reduction of approximately 25.7\% in NOx and 14.4\% in carbon dioxide (CO2) emissions. Nevertheless, there remains a notable lack of knowledge regarding the estimation of pollutant concentration reduction when vehicles with electric engines are introduced either entirely or partially. Realizing substantial air quality improvement necessitates the widespread adoption of electric vehicles, encompassing both light cars and heavy-duty vehicles \cite{ferrero2016impact}.

Given the fluctuating electricity generation mix, it is imperative to scrutinize the charging habits of BEV users. Analyzing and, when necessary, controlling these habits can minimize emissions and costs. Notably, recharging vehicles at home can increase emissions from the grid, especially during coincidences with afternoon peaks when drivers arrive home. This increased peak load may necessitate additional generating capacity, leading to higher costs and emissions associated with plug-in vehicles.\cite{sioshansi2010cost}

Efforts to smooth and ideally flatten the aggregate demand through the strategic shifting of loads in time can assist utilities in avoiding blackouts or brownouts and maintaining a reliable service.\cite{thornburg2017simulating} While Electric Vehicles (EVs) can function as shiftable loads or energy storage for the power grid, most utilities have traditionally treated them as simple loads. Uncoordinated charging, where EVs start charging immediately upon being plugged in, adds to the demand curve and poses a burden during high-demand times. However, with controlled and coordinated charging aligned with grid needs, EVs can serve as both flexible loads and energy storage. Without such coordination, the widespread adoption of EVs could potentially create supply challenges for the electric power grid at scale.\cite{Mohammadi2023}\cite{verzijlbergh2012network}

Looking beyond energy generation and consumption, the type of EV becomes a crucial factor influencing the environmental benefits of electric mobility. EVs are categorized into four main types: Hybrid Electric Vehicles (HEVs), primarily powered by gasoline with a small battery supporting the combustion engine; Plug-in Hybrid Electric Vehicles (PHEVs), powered independently by both gasoline and electricity; Battery Electric Vehicles (BEVs), solely powered by electricity; and Fuel Cell Electric Vehicles (FCEVs), powered by hydrogen.\cite{requia2018clean}


\section{Background}


In recent years, the electricity network has undergone a significant transformation driven by the increasing penetration of renewable energy (RE) generation and the modernization of transport infrastructure. Policymakers, engineers, and business leaders are actively seeking alternative energy solutions that are both economically viable and environmentally friendly, given strains on petroleum reserves and the pressing issue of GHG emissions. RE and EVs emerge as promising answers to provide energy cost savings and emission reductions \cite{kempton2009test}. In Texas specifically, the generation mix includes multiple RE. While biomass and hydro make modest contributions, nuclear power emerges as a significant player. Solar energy is substantial, and wind power stands out prominently, reflecting the states favorable conditions for wind energy generation and the investments therein. This mix of renewable sources showcases a forward-thinking approach, contributing not only to environmental sustainability but also ensuring a resilient and diversified energy portfolio.

\begin{figure}[h]
    \centering
    \includegraphics[width = .5\textwidth]{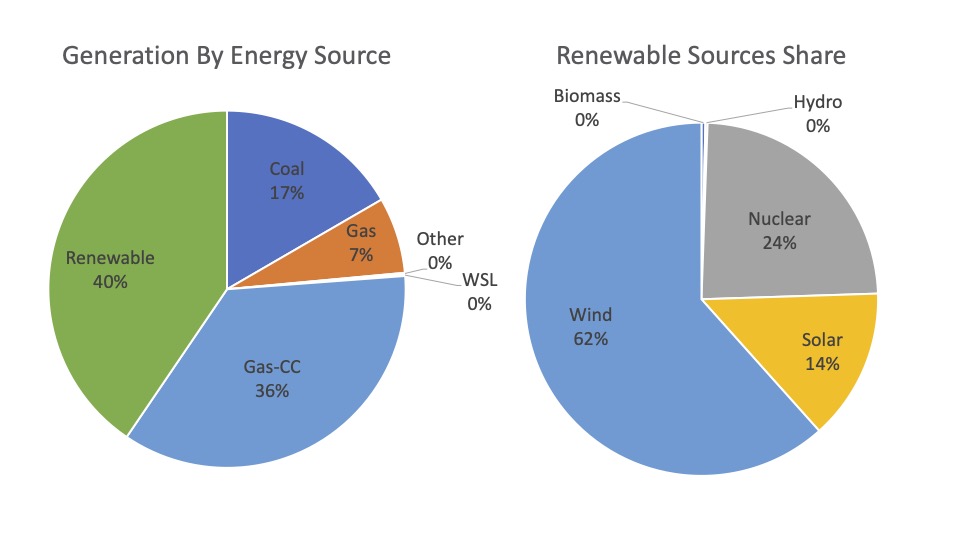}
    \caption{Power Generation Mix in Texas, USA (2022)}
    \label{fig:VPP}
\end{figure}

Integrating RE sources into the electric grid presents challenges due to the intermittent nature of generation sources and their inconsistency with energy usage \cite{jin2014optimized}\cite{loutan2007integration}. This complexity is a potential hurdle in transitioning to a more sustainable energy landscape. Moreover, the electric power generation and transportation sectors are major contributors to petroleum shortages and GHG emissions \cite{kempton2009test}.

\subsection{Smart Grid}

To overcome these challenges for a greener future, the concept of the smart grid has gained prominence. This next-generation power grid employs bidirectional power flow and a robust communication network to create a globally distributed and integrated energy distribution network. The smart grid can align power networks with environmental targets, accommodate EV adoption, and facilitate distributed energy generation with storage capabilities \cite{ram2021review}.

However, integrating EVs into the electric distribution network poses stability concerns, particularly with the increased penetration of EV charging. The charging patterns, as revealed by a study from the National Renewable Energy Laboratory (NREL), coincide with the grid's peak load demand profile, potentially leading to higher costs for EV owners during peak hours \cite{gilleran2021impact}.

An EV’s charging rate (i.e., the speed of charging the EV battery) is controllable by the EV owner, the grid, or both depending on the equipment. This controllability positions EVs as not only consumers but also as controllable loads in grid systems, allowing EVs to serve as distributed energy storage when charging is managed strategically and vehicle-to-grid (V2G) technology is in place \cite{Mohammadi2023}.

The power grid faces challenges in meeting the rapid surge in electricity demand, emphasizing the need for higher reliability and efficiency. Current power grids are designed to cater to peak demands, resulting in underutilization and waste of natural resources. Fast-responding generators, such as those that burn fossil-fuels, come with significant costs and a substantial GHG footprint.

In response to these challenges, the evolution of the smart grid, coupled with the integration of RE sources, advanced metering infrastructure, smart meters, EVs, and dynamic pricing, has gained momentum \cite{tushar2017demand}. Additionally, the widespread deployment of home energy management systems (HEMS) and communicating devices is expected to upgrade the existing power grid, transforming it into a more intelligent and decentralized system \cite{caron2010incentive}. These advancements collectively contribute to addressing the demand-side management (DSM) problem efficiently, marking a significant stride toward a sustainable and resilient energy future \cite{tushar2017demand}.

\subsection{Virtual Power Plants}\label{sec:scalable}

The advent of Virtual Power Plants (VPPs) presents a transformative opportunity to alleviate the load on power networks. By generating power locally and sharing it among participants without the need for long-distance transmission at high tension, energy loss factors are either minimized or eliminated. This marks a significant shift in energy dynamics, where participants are no longer passive users but active influencers within the power system, albeit within defined limits—participants are not tasked with directly controlling devices' on-off switches.

At the helm of the typical VPP is a computer system overseen by the Distribution System Operator (DSO), potentially organized on the foundation of an artificial neural network. Remarkably, even a household with as little as 1 kW of distributed generation capacity (be it PV, FC, or combined heat and power, for example) can support a VPP. The key lies in connecting all these sources and orchestrating their operation to achieve a state of self-balance in the most effective manner. The VPP emphasizes local generation, enabling central generation to operate under more stable conditions. DSOs can optimize peaks in heat and power demands more efficiently, with the incorporation of storage for heat or electricity further enhancing the VPP's operational conditions \cite{nikonowicz2012virtual}. An example VPP configuration is shown in Figure \ref{fig:VPP}.

\begin{figure}[h]
    \centering
    \includegraphics[width = .6\textwidth]{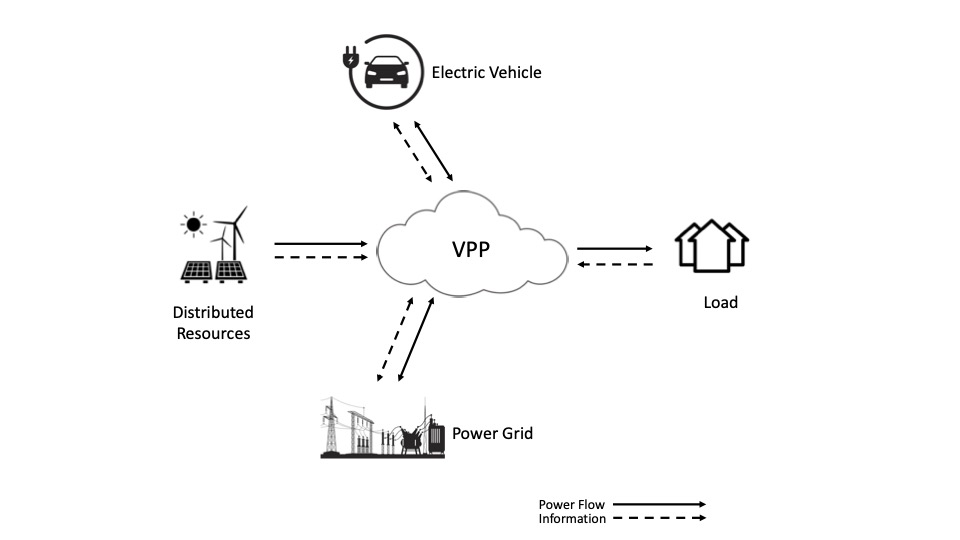}
    \caption{Example structure of a virtual power plant}
    \label{fig:VPP}
\end{figure}

\subsection{Energy Storage Systems and Distributed Energy Resources}

As sporadic renewable generation increases, an energy storage system (ESS) become crucial for ensuring stable, reliable, and consistent EV charging throughout the day \cite{ram2021review}. The ESS is a vital tool for adapting power demand variations to the level of power generation in real time, particularly given the weather-related intermittency of RE. These systems can serve as additional sources or energy buffers for non-dispatchable or stochastic generation, such as wind turbines or PV technologies, especially in weak networks \cite{saboori2011virtual}.

EVs are evolving into distributed energy resources (DERs) that are poised to play a pivotal role in future smart power networks. What sets EVs apart from other DERs is their mobility, allowing them to connect at different points in the grid while maintaining consistent service quality.

VPP generation primarily relies on RE resources, whose production is challenging to perfectly forecast especially on a day-ahead schedule. In addition to storage units, EVs can contribute to balancing forecast errors by leveraging their capacity capabilities. Moreover, the charging of EVs can be strategically implemented during periods of high solar and wind generation \cite{raab2011virtual}.

Fleet owners typically charging their EVs with a negotiated flat electricity tariff, but they can unlock additional benefits by actively participating in the wholesale market. This market presence requires fleet owners to determine the quantities of electricity they are willing to store and sell back to the grid, along with setting minimum and maximum prices for each time slot one day ahead. These parameters are then matched with those of other buyers and sellers in the day-ahead electricity wholesale market \cite{kahlen2018electric}.

However, managing a multitude of DERs under market conditions introduces new challenges. Weather-dependent DER technologies, like solar cells and wind turbines, present fluctuating and non-dispatchable outputs, limiting their contribution and incurring economic penalties. Additionally, isolated operation due to diverse ownership hampers cooperation and communication between neighboring DER units. Aggregating these units into a Virtual Power Plant (VPP) is a strategic solution to address these issues, promoting synergy and enhancing the capability of DERs to serve not only local needs but those of the entire grid \cite{saboori2011virtual}.


\section{Methods}
The integration of electric vehicles (EVs) into future smart grids is crucial for successful energy transition. Under the Virtual Power Plant (VPP) concept, EVs can be viewed as moving battery storages, contributing to the efficiency of integrating renewable sources with their alternating generation. Managing EV charging is essential from various perspectives, including energy resilience and grid operations, as studies indicate a significant impact on the wholesale electricity market.\cite{foley2013impacts}

When EVs are left uncoordinated in their charging, it poses challenges for the electric power grid. This is particularly evident when numerous EVs plug into chargers simultaneously, such as commuters arriving home after work, or when many EVs disconnect from chargers simultaneously. Unmanaged, these scenarios can lead to spikes or sharp drops in power demand, respectively. To address this, centralized or decentralized approaches, especially demand response (DR) strategies, can smooth demand shifts and reduce strain on the grid's existing supply and storage infrastructure. Therefore, effective EV and grid interactions are essential for a successful energy transition.\cite{Mohammadi2023}

The use of machine learning (ML) algorithms, based on historical data of charging load and user behavior, offers a valuable tool for training and learning trends and patterns. After the training phase, accurate predictions can be made, enhancing EV charging scheduling strategies. ML algorithms have proven capable of providing good forecasts for timeseries data, making them suitable for charging behavior predictions. These predictions, whether used independently or in conjunction with other algorithms, play a crucial role in optimizing EV charging and mitigating potential impacts on the electricity market.\cite{shahriar2020machine, MohammadBOOKCHAPTER}

In this paper, two approaches were employed to determine the optimal charging time for electric vehicles (EVs) in Texas, considering the grid situation. One approach is based on the grid load, and the other approach incorporates both grid load and renewable sources generation. Detecting the trend of electricity consumption throughout the day is crucial for identifying the optimal charging time. Recognizing that one trend does not sufficiently characterize the entire year, the days were grouped based on similarity, and trends were identified for each group. The data set includes hourly load and renewable generation data for 2022.

To analyze load and renewable sources trends on different days, this research utilized the dynamic time warping (DTW) clustering model. The Electric Reliability Council of Texas (ERCOT) provided time-series data, capturing load and renewable generation in Texas. Measuring similarity in time-series data is challenging due to the need to consider the order of elements in the sequences. The DTW algorithm was chosen for its ability to measure the similarity between two temporal sequences, allowing for nonlinear warping in the time dimension to determine their similarity, irrespective of certain nonlinear variations in time \cite{Mohammadi2023}\cite{yao2021clustering}.

DTW clustering is chosen for the following advantages over other methods, and it contrasts with traditional approaches like Euclidean distance. While Euclidean distance is suitable for spatial calculations, DTW excels in capturing temporal differences in time-series data. Assessing time series similarity requires a method that leverages both temporal and spatial characteristics \cite{wang2018time}.
To further analyze the trends identified for each cluster, thresholds were needed to determine the best time for charging and the optimal potential time for vehicle-to-grid (V2G) interactions. The K-means clustering model, an unsupervised machine learning approach, was employed to find these thresholds. K-means clustering attempts to group comparable observations by minimizing the Euclidean distance to centroids, providing valuable insights for optimizing EV charging and V2G interactions \cite{abo2022optimal}.

 \section{Results}

\begin{figure*}[t]
    \centering
    \includegraphics[width=5.3in]{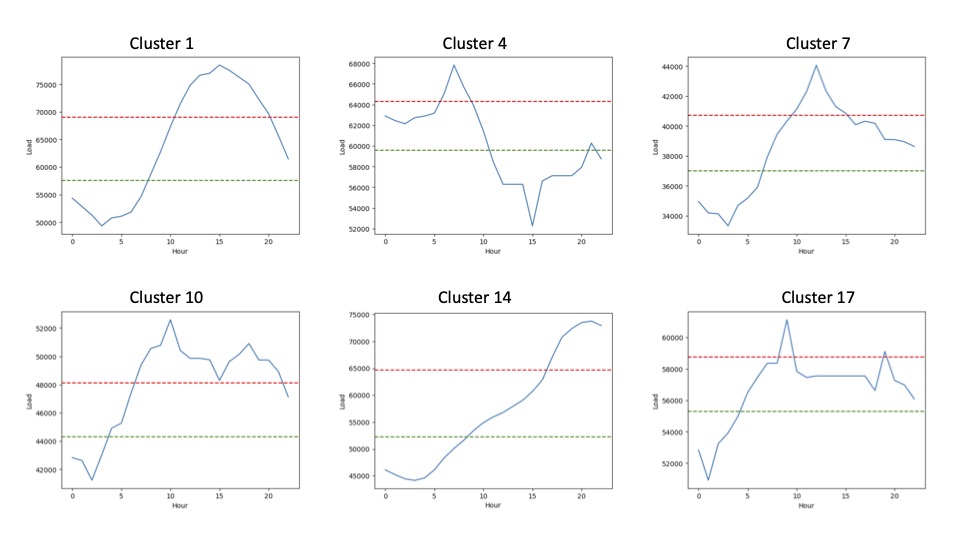}
    \caption{Representative clusters of twenty cluster patterns  in hourly load trend}
    \label{fig:Load}
\end{figure*}

In this paper, we employed a robust methodology that leverages a comprehensive dataset encompassing the total load and renewable generation data for each hour throughout the entirety of 2022 in the state of Texas. The focal point of this analysis lies in the application of DTW clustering, a technique adept at discerning temporal patterns in sequences. Specifically, we utilized DTW clustering to categorize days into distinct clusters based on two pivotal metrics: total load and net load. The total load represents the aggregate electricity consumption in Texas for each hourly interval, while net load is derived by subtracting the total renewable generation from the total load. This innovative approach allows for a nuanced understanding of how daily electricity consumption and renewable energy generation vary over the course of the year.

Moreover, this approach goes beyond simple classification; it explores the identification of prevalent trends within each group. This stage is essential for gaining insights into the overarching patterns that define different periods throughout the year. To enhance the practicality of the results, we integrated k-means clustering to identify two thresholds within each cluster. These thresholds, revealed through k-means clustering, act as crucial markers to determine the optimal timing for both charging and V2G activities. Through the utilization of a dual clustering method—DTW for classification and k-means for threshold identification—we've developed a sophisticated methodology. This methodology not only categorizes days into meaningful groups but also extracts actionable information for strategic decision-making in the realm of energy consumption and renewable integration.

\subsection{Total Load}

We employed the total load data for each hour throughout the entirety of 2022 in Texas to categorize days into 20 distinct clusters. Through extensive experimentation, our study determined that 20 clusters represent an optimal grouping for our analysis. For each cluster, a representative trend was identified, utilizing k-means clustering to establish two crucial thresholds represented by green and red horizontal lines. In Figure \ref{fig:Load}, we showcase six illustrative examples of these clusters, emphasizing the varied load ranges observed along the vertical axis. When the trend falls below the green line within a cluster, it designates off-peak times, suggesting this range of hours as ideal for EV charging. Conversely, when the trend exceeds the red line, it signifies peak times, advising against EV charging while presenting an opportunity for V2G activities during these intervals. This approach allows for a clear identification of optimal charging and V2G windows tailored to the specific load characteristics of each cluster.

\subsection{Net Load}

In the Net Load approach, we calculate the net load by subtracting the renewable generation from the total load, representing the demand supported by non-renewable sources. This method offers a distinct advantage over the total load approach, particularly when considering EV charging. The potential issue with relying solely on the total load is that the peaks of renewable generation and load may coincide, alleviating pressure on nonrenewable sources and making V2G less crucial. To address this, the net load approach accounts for the demand supported by non-renewable sources, providing a more nuanced perspective.

\begin{figure*}[!ht]
    \centering
    \includegraphics[width=5in]{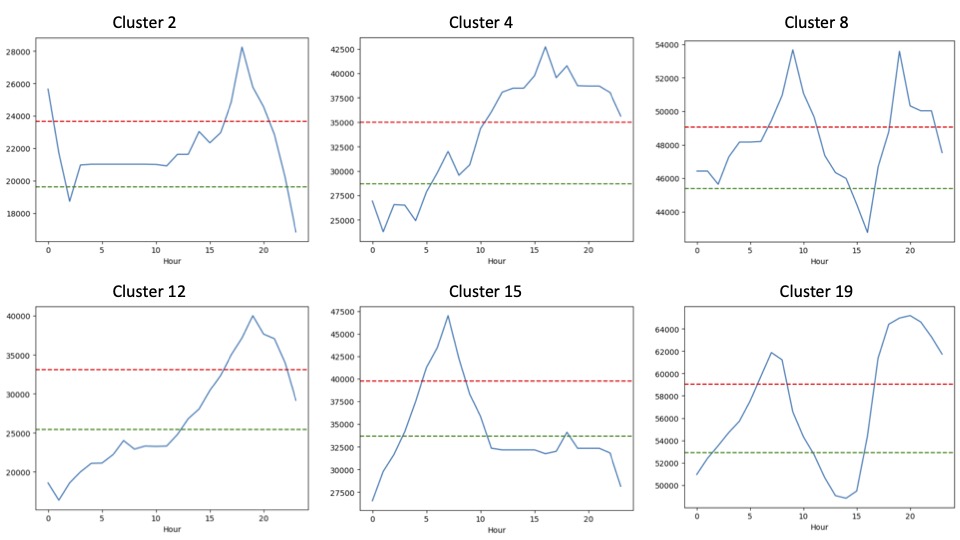}
    \caption{Representative clusters of twenty cluster patterns in hourly net load trend}
    \label{fig:load-renewable}
\end{figure*}

To implement this approach, we utilize 20 clusters as determined to be optimal for our analysis. For each cluster, a representative trend is established, and two thresholds are identified to demarcate crucial ranges. Figure \ref{fig:load-renewable} illustrates six exemplar clusters, where ranges above the red line signify instances of high load and low renewable generation. Leveraging V2G for EVs as a virtual power plant during these periods can effectively alleviate stress on the power grid, promoting a more environmentally friendly power generation by optimizing the utilization of renewable energy sources. This approach thus enables targeted recommendations for V2G implementation during specific load and renewable generation scenarios within each identified cluster.

\section{Conclusions and Future Work}
Our exploration into the connection between EVs, smart grids, and renewable energy in Texas illuminates key pathways for sustainable energy transitions. This study reveals the intricate relationship between the charging patterns of electric vehicles and the dynamics of the grid. It emphasizes the significance of strategic interventions to optimize energy utilization. Leveraging DTW clustering and k-means methodologies, we not only categorized daily energy consumption and renewable generation but also identified tailored windows for EV charging and V2G interactions. This understanding contributes actionable insights for users and grid operators alike, paving the way for a resilient, environmentally conscious energy future. Moving forward, collaborative efforts and policy measures must align to harness the full potential of EVs within smart grids, emphasizing predictive scheduling, VPPs, and a collective commitment to sustainable energy practices. This study guiding the integration of EVs into the smart grids and leading us toward a future where clean energy and grid stability are connected.

The work described is being developed and tested by Grid Fruit, LLC. Grid Fruit is a private software company that puts unused energy and operational data to work for the grid and its customers \cite{mohammadi2020federated}.


\bibliographystyle{./bibliography/IEEEtran}
\bibliography{./bibliography/IEEEabrv,./bibliography/IEEEexample, ./bibliography/refs}
\end{document}